\documentclass[10pt,twocolumn,english]{IEEEtran}
\usepackage[T1]{fontenc}
\usepackage[a4paper]{geometry}
\usepackage{array}
\usepackage{graphicx}
\usepackage{setspace}
\PassOptionsToPackage{normalem}{ulem}
\usepackage{ulem}
\usepackage{nomencl}

\providecommand{\makenomenclature}{\makeglossary}
\makenomenclature

\makeatletter

\newcommand{\noun}[1]{\textsc{#1}}
\providecommand{\tabularnewline}{\\}

\usepackage{nopageno}
\makeatother

\usepackage{babel}
\begin{document}
\begin{singlespace}

\title{A Discussion on Developing Multihop Routing Metrics Sensitive to
Node Mobility}
\end{singlespace}

\author{Namusale Chama{*}, Rute Sofia{*}\thanks{*(namusale.chama,rute.sofia)@ulusofona.pt. Internet Architectures and Networking (IAN), Research Unit in Informatics Systems and Technologies (SITI), Universidade Lusófona. Campo Grande 376, 1749-024 Lisboa, Portugal.
** The research leading to these results has received funding from the EU IST Seventh Framework Programme ([FP7/2007-2013]) under grant agreement number 257418, project ULOOP  User-centric Wireless Local Loop.}
\\
SITI, University Lusófona }

\maketitle
\thispagestyle{empty}
\begin{abstract}
This paper is focused on a discussion of parameters and heuristics
that are expected to assist multihop routing in becoming more sensitive
to node mobility. We provide a discussion concerning existing and
a few proposed parameters. Moreover, the work also discusses two new
heuristics based on the notion of link duration. The heuristics are
compared based on a meaningful set of scenarios that attain different
mobility aspects.\end{abstract}


\section{Introduction}

\begin{singlespace}
The most recent paradigms in wireless architectures describe environments
where nodes present a somewhat dynamic behavior (e.g. \emph{Mobile
Ad-hoc Networks}, \emph{MANETS}) or even a highly dynamic behavior
(e.g., \emph{User-provided Networks}, \emph{UPNs}). Nodes in such
environments correspond to wireless devices which are carried or controlled
by humans and hence exhibit movement patterns which mimic the ones
of humans - \emph{social mobility patterns}. Moreover, in such environments,
data transmission is based on multihop routing, namely, on \emph{single-source
shortest-path} approaches. In terms of routing metrics, the most popular
multihop routing approaches rely on static link cost metrics such
as hop count. The result is that when facing movement of nodes, multihop
routing has its own shortfalls e.g., the need to recompute paths frequently
if nodes exhibit high variability in movement of the nodes. In other
words, current multihop approaches lack sensitivity in what concerns
nodes movement. The routing metrics currently being considered cannot
capture node speed variation or acceleration, node movement pattern,
or even direction. Hence, a topology change is merely interpreted
as a trigger to perform path re-computation. There are, however, cases
where node movement may actually not represent a link break. Or, instead,
a link change due to node movement may be so subtle that in fact it
would not require any update to the topology, and hence, current approaches
may result in useless path re-computation, of which the cost relates
to additional signaling overhead and latency.
\end{singlespace}

This paper is focused on a brief analysis of mobility impact on routing
and proposes a category of routing metrics related to different notions
of link duration. Our expectations are that such a metric may make
multihop routing more sensitive to node movement and hence, assist
in improving the trade-off of node movement vs. network efficiency.
Under such category, we provide several metrics and a discussion on
the impact that they have, under different parameters.

The remainder sections are organized as follows. Section \ref{sec:RelatedWork}
goes over work that is related to ours, highlighting the relation
between previous work, and our contribution. Section \ref{sec:Impact}
gives an overview on the notion of node movement and how it impacts
routing and the definition of a wireless link. Section \ref{sub:Parameters}
gives an overview of the existing mobility tracking parameters. Section
\ref{sec:Making-Routing-Mobility} describes our proposed routing
metrics and how they improve routing sensitivity to node movement,
while section \ref{sec:Analysis-1} provides an initial performance
comparison of the proposed metrics. We conclude in section \ref{sec:Concl}.

\section{Related Work}

\label{sec:RelatedWork}

A number of approaches have been dealing with detection and measurement
of accurate node mobility as well as counterbalancing mobility impact
on routing. 

A first category of related work considers applying signal strength
measurement at the receiver as a way to estimate the distance variation
between two nodes, having as ultimate goal providing a way to build
more robust paths. For instance, the \emph{Mobility Prediction Routing
Protocol (MAODV)}\cite{mengmobility} relies on the variation of the
received signal strength by a particular node to predict link breaks.
Dube et al. \cite{dube_ss} have also applied signal strength measurement
as a way to build more robust paths. Manoj et al. applied the notion
of received signal strength to track the distance variation between
two nodes, in a specific time period \cite{Manoj_LinkLife}. Their
distance change tracking approach is a desirable feature as it can
capture movement of nodes and the impact on links. However, and similarly
to the other work in this category, it does not track movement patterns. 

A second category of work that tries to make routing more sensitive
to node mobility relates to throughput variation measurement as a
way to determine node mobility. For instance, Suyang and Evans have
used the slope of change of throughput in a link vs. the link load
to estimate topology changes \cite{suyang_ju_throughput}. Based on
history, through throughput monitoring, a decrease, according to Suyang
and Evans, means that there is a true change in the physical topology.
They have attributed the changes to increase in the node distance
and increase in interference. Even with good attributes of avoiding
interference and detecting mobility collectively, node mobility individually
has attributes that have not been addressed. Nodes may exhibit movement
that does not necessarily impact the route stability.

A third category of work that we cite relates to an attempt that is
not directly tackling mobility impact on routing in the sense of reducing
computation but instead addresses a latency reduction. In such category
falls the \emph{On-Demand Multipath Distance Vector} (\emph{AOMDV})
\cite{marina_AODV} and the work of Kim et al. \cite{kim_multipath_disjoint}
which considers disjoint paths in AODV to counter link failure in
case the alternative path shares some links with the primary one.
Albeit multipath assists in reducing latency and packet loss in the
event of link failure, the cost of path re-computation is still present
and affects the network operation.

A fourth category of work relies on \emph{link sensing} as a measure
of improving routing in terms of mobility sensitivity. Benzaid et
al. \cite{benzaid_fastolsr} have proposed \emph{the fast Optimized
Link State Routing Protocol (Fast-OLSR)} whose basic idea is to detect
link changes in a quicker way, by increasing the HELLO sending rate.
Albeit interesting, such rate only assists in understanding that some
nodes may be on the move, but not exactly which.

All of the mentioned approaches have in common the aspect that they
fail in being able to distinguish between a topology change that is
long-lasting, and a topology change that is so short that in fact
it should not result in route re-computation. Our belief is that by
defining a multihop routing metric more sensitive to node mobility,
multihop routing can become more robust and better adjusted to the
current wireless dynamic scenarios.

\section{Node Mobility Impact on Routing }

\label{sec:Impact}

This section provides a brief discussion on the impact of node mobility
on routing. \emph{Node mobility} here refers to a change in speed
(inclusive of relative speed) and/or direction (also relative direction)
for a node. Hence, \emph{node movement} is based on three main aspects:
node position; speed; direction. 

The impact of mobility on routing can be measured mostly by analyzing
the trade-off in robustness (e.g. the need to recompute paths more)
vs. signaling overhead (more messages sent to quicker detect link
breaks). Furthermore, node mobility impacts routing on a different
number of ways and here we shall address the main ones, namely, relation
to distance; movement pattern (and how it affects links); relative
movement (link remains stable due to similar movement of the nodes
that compose the link); impact on the different stages of the routing
process (e.g. route discovery and maintenance phase). Intuitively,
it seems that node movement can be better captured just by sending
additional signaling messages. However, it is not always the case
that by increasing signaling overhead one may better prevent link
breaks and reduce the cost of re-computation; on the other hand, node
movement patterns affect links temporarily or permanently. 

\begin{figure}[b]
\center\includegraphics[scale=0.4,bb = 0 0 200 100, draft, type=eps]{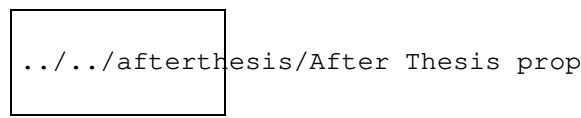}

\protect\caption{Examples of node mobility.\label{fig:Node-Mobility-and-Routing-1}}
\end{figure}

To give a concrete example let us consider Figure \ref{fig:Node-Mobility-and-Routing-1}
which illustrates a wireless topology where A and B represent nodes
in movement. The figure considers three different cases. In Figure
\ref{fig:Node-Mobility-and-Routing-1} i), A exhibits confined movement
eventually returning to its original position. In ii) B is the node
moving in a \emph{ping-pong} pattern, i.e. B is jumping back and forth
between two different positions. This stands for a case of repetitive
movement, where the node exhibits some pattern and frequency of moving
away/returning to origin. The final case (cf. Figure \ref{fig:Node-Mobility-and-Routing-1}
iii)) corresponds to the case where B permanently moves away from
its original position. Upon movement of at least one of the nodes
A and B, the corresponding link quality is affected. If the nodes
exhibit frequent movement, frequent path re-computation may occur.
For scenarios such as the one illustrated in Figure \ref{fig:Node-Mobility-and-Routing-1},
it may happen that two neighbor nodes move away from each other resulting
in a link being broken and consequently, resulting in topology re-computation,
to return to their original position a few milliseconds later. Were
it for a protocol truly adaptive to node movement, the decision on
whether or not to trigger path re-computation should be based on metrics
that can capture node mobility patterns. 

Node mobility impact on routing is also distance dependent, i.e.,
related to link\emph{ size}. For instance, a link formed by two nodes
far apart (\emph{long link}) can be affected and broken even by a
small, insignificant node movement. If instead we have a \emph{short
link} (small distance between the two nodes), the movement of a node
has to be significant to result in a link break. It should be noticed
that link quality may degrade also due to mobility, but what we are
highlighting here is that changes in distance are not sufficient to
define a routing metric sensitive enough to node movement. It is also
necessary to incorporate some sensitivity to a node's movement pattern
and this is not a trivial task given the possible mobility patterns.
For instance, a node moving between two different positions A and
B can just move from A to B; move away from A to B and come back to
A; or it can be ping-ponging between A and B. As mentioned previously,
such movement pattern may be insignificant in terms of impact on a
link (e.g. because the distance between the nodes is short). In contrast,
a move between A and B will impact link capacity heavily, and a ping-pong
movement will result in a wide wireless link capacity variation. For
both the aforementioned cases, a link is not truly broken and yet,
today path re-computation would occur.

Another aspect to consider in terms of impact of mobility on routing
is the routing phase where movement is detected. Of utmost relevance
is node mobility during the route discovery and route maintenance
phases. In relation to mobility impact, the most relevant routing
processes are: \emph{route discovery}, where a route to a particular
destination is not known and has to be built and computed and \emph{route
maintenance,} where routes are maintained and recomputed. For instance,
if a link on an active route breaks, an alternative route to the destination
is computed and this is done during the the route maintenance phase.
Assuming the existence of a link considered during route discovery,
where one of the nodes is moving, it may happen that the resultant
path of route discovery is not available as the link would have been
broken. Routing protocol under route maintenance will have to recompute
an alternative path for data transfer.

To further debate on the impact of node mobility has on the routing
process, we discuss, in the next sections, impact of movement during
route discovery and maintenance. We start by addressing such potential
impact by analyzing different parameters, to then briefly the potential
impact on the two most popular multihop routing families distance
vector and link state, which are represented here, by the \emph{Ad
hoc On-Demand Distance Vector (AODV) }\cite{royer_aodv} and the \emph{Optimized
Link State Routing Protocol (OLSR)}\cite{Jacq_OLSR},\emph{ }respectively.
The explanation provided next has as main purpose to explain in further
detail the impact of node movement on routing.

\subsection{Movement Impact during Route Discovery}

Considering AODV, during the route discovery, a node (i.e. source
node) upon demand broadcasts \emph{Route Request (RREQ)} packets.
An answer in the form of a \emph{Route Reply} (RREP) will be returned
as soon as a node realizes it has a route established to the destination.
Assuming node movement of the source node, or movement of any of the
nodes on the would be path before an answer is returned, i.e., before
a path is fully established. This may result in path establishment
failure, depending on the type and pattern of movement of the source
node and the intermediate node. To what extent node mobility will
impact the route discovery phase depends on several aspects, for instance:
whether a link is short or long, the mobility pattern, frequency of
motion and also the node degree. The number of nodes moving as well
as the sequence of node displacement from the original positions is
also relevant to address.

\begin{figure}[b]
\includegraphics[scale=0.7,bb = 0 0 200 100, draft, type=eps]{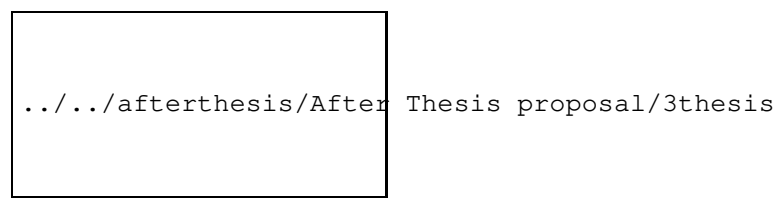}

\protect\caption{Node Topology Example.\label{fig:Node-Topology-Example}}
\end{figure}

To assist the explanation the impact of node mobility on route discovery,
Figure \ref{fig:Node-Topology-Example} shows a topology where a route
has to be discovered from Node S to node D, where, based on hop count,
the path between S and D should be S-C--D a number of alternative
paths also exist but are longer, for examplepaths S-A-B-D,S-A-B-C-D
and S-E-F-G-D to mention a few. Let us consider that nodes A and B
exhibits some form of confined movement as shown in Figure \ref{fig:Node-Mobility-and-Routing-1}
i) and node S broadcasts a RREQ for route discovery. Let us also assume
that only C and G know about the whereabouts of D. The RREQ sent by
S may, due to such confined movement, reach C later than it reaches
G. Therefore, the answer in the form of a RREP may result in a route
that may be formed earlier may actually be the longest and the shortest
may appear to be longer, for example, having a path of S-A-B-C-D instead
of S-C-D. 

The impact of movement in this phase is therefore related to the type
of movement but also related to the type (short or long) link. If
the links affected by confined movement are long, then the node movement
will affect more significantly the path recomputation and adequate
route discovery. If links are short, then movement of nodes in a confined
area may not even be noticeable from a routing process perspective.
However, if we consider some movement pattern which exhibits some
regularity, such as the one in Figure \ref{fig:Node-Mobility-and-Routing-1}
ii), then the frequency of regular movements significantly affects
the route discovery independently of links being short or long.

We have discussed the impact of confined node mobility on route discovery,
now let us debate on the frequency of regular node movement and how,
in our opinion, it impacts the routing process in terms of route discovery.
The three cases of node movement presented in Figure \ref{fig:Node-Mobility-and-Routing-1}
may exhibit some frequency which implies that the node crosses its
original position at some instant in time. By \emph{low frequency}
of movement is here meant that the node crosses its original position
seldom; by \emph{high frequency} it is meant that the node crosses
its original position often. 

Assuming a confined movement scenario for a node with low frequency
of movement route discovery is barely affected with such node mobility
whether long link or short links,while high frequency of confined
movement impacts more on long links than short links. However, if
we consider a movement pattern such as a ping-pong movement, then
the impact of such frequency may in fact severely affect the route
discovery phase, leading to route that are not shortest-path based
or even delaying such phase in a significant way as both short and
long links are affected.

Such delay is highly related to the network composition and node degree,
in particular on the degree of the nodes exhibiting movement. Moreover
it is also highly related to the relation of the movement that a node
exhibits in regards to its neighbors. Therefore, it is not always
the case that links formed with a node that exhibit high frequency
of movement will significantly impact the route discovery phase. Such
impact also depends on the relativeness of movement between nodes
that form links. To give an illustration, we refer to Figure \ref{fig:Node-Topology-Example}.
If all nodes in this topology were moving with high frequency of ping
pong with minimal relative movements among the nodes, delay in route
discovery will be minimal. On the other hand, high variability of
such movement at high frequency will introduce high delays in the
route discovery stage.

In terms of relation to distance-vector and during the route discovery
phase, the main impact of node movement is delay which in our opinion
may, in specific cases and due to node frequency as well as topology
composition, result in such variability that mainly jeopardize the
whole routing process, as the route discovery phase may be delayed
significantly for a distance-vector approach. 

Were it for the case of a link-state protocol such as OLSR, then route
discovery is performed in a proactive way based upon the Hello control
messages and also based on the notion of \emph{Multipoint Relay (MPR)},
i.e., a successor of a node chosen to forward packets on behalf of
the node. This assists in reducing flooding on the network, in contrast
to the original link-state routing approaches. OLSR is better suited
for large, dense environments. Moreover, OLSR can tune the frequency
of information exchange and thus provides, in principle, better support
for node movement.

In terms of the route discovery phase, the impact that node movement
again relates to the parameters already discussed, namely: correlation
of movement pattern to time (frequency and type of movement) as well
as to node degree and network density. The main difference in comparison
to AODV during this phase relates to the proactive behavior, which
provides more stability when nodes move. Let us again consider Figure
\ref{fig:Node-Topology-Example} and the provided (potential) paths.
Again assuming that nodes A and B exhibit some type of movement, the
expected delay would most likely be less but the result would be the
same in the sense that again the selected route would be the longest
one independently of the fact that the movement type and frequency
could imply that after a small delay the best route could indeed be
the shortest one. In other words: none of these families currently
includes a natural (metric-based) way to detect such minor variations
and to ignore them. 

In terms of different multihop approaches, while with AODV the result
may be a significant delay while with OLSR the delay may be smaller
but the signaling overhead may significantly increase, depending on
the type of movement, frequency of movement, as well as related to
the position of the node(s) moving from and end-to-end path perspective.

\subsection{Movement Impact during Route Maintenance }

Again considering the AODV perspective, let us now debate the potential
impact that node movement has during the route maintenance phase.
A topology change may occur due to a temporary link break, or due
to a permanent link break. For instance, if one node moves from a
specific position to another and hence there is a link break but the
node returns to its original position in a few milliseconds, this
corresponds to a \emph{temporary} link break. A temporary link break
does not always imply discontinuity from a routing perspective and
this is highly related to the type and pattern of movement. For instance,
if a node exhibits confined movement as discussed the frequency of
movement dictates whether or not such movement may result into a temporary
link break, or a permanent link break. Being on-demand (reactive),
upon even a temporary link break AODV triggers signaling in order
to deal with topology changes. It may even happen that upon the detection
of a temporary break, AODV triggers path recomputation and the result
may simply be the path that was already established. This will increase
the signaling overhead in a way that could be prevented, if the applied
routing metric would be capable of ``isolating'' these situations,
i.e., by making the routing protocol understand when a change is temporary,
or permanent, or simply react for cases where changes are permanent.

Assuming that the type of movement implies some frequency of returning
to the original position, then in addition to the signaling overhead
there is delay which is highly dependent on such frequency.

In regards to the impact on a link-state approach (e.g. OLSR) during
the route maintenance phase, due to its proactive nature, OLSR will
detect quicker the location of a topology change and due to the flooding
nature, it will, most likely, heal such failure quick. However, for
the case of a temporary failure there is no detection capability.
Both temporary and permanent topology changes will be dealt with as
a change and hence require recomputation. Signaling overhead is associated
with this. The corresponding delay will be less for OLSR than for
AODV. However, both families treat temporary and permanent topology
changes as changes, thus requiring path recomputation. Despite the
fact that such changes may be insignificant, the relative cost (be
it in terms of delay or in terms of signaling overhead) seems to impact
both protocol families the same way. This has, of course, to be corroborated
with experimentation which we leave as future work.

Therefore, one main aspect to tackle in order to make routing more
sensitive to node mobility is to consider metrics that are capable
of capturing some properties of such movement. In the next section,
we describe a number of parameters and of metrics that can be used
to achieve such goals.

\section{Mobility tracking Parameters}

\label{sub:Parameters}

\begin{singlespace}
In this section, we describe several parameters related to the routing
process which provide some support in terms of sensitivity to node
mobility, explaining their advantages, as well as weaknesses.
\end{singlespace}

A number of parameters affect the performance of routing protocols
differently. They will affect the routing protocol on the routing
discovery and route maintenance procedures. The parameters under study
in this section are: \emph{link duration} or \emph{lifetime}; \emph{node
degree stability}; \emph{ratio of static nodes vs. moving nodes};
\emph{average number of link breaks}; \emph{pause time}.

\subsection{Link Duration}

\emph{Link duration (LD)\nomenclature{LD}{Link Duration}} or \emph{lifetime}
is a parameter that is tightly related to the movement of nodes and
is also as of today one of the parameters that is most popular in
terms of tracking node mobility. By definition, link duration is associated
to the period of time where two nodes are within the transmission
range of each other. In other words, it is the time period that starts
when two nodes move to the transmission range of each other and that
ends when the signal strength perceived by the receiver node goes
under a specific threshold \cite{liangLD}\cite{wu_LD} \cite{tsao_LD}.
Some authors then provide a variation of this definition by working
the threshold value.

Today's definition of LD only assimilates node mobility in regards
to its relation to signal strength. It fails, however, in terms of
sensitivity to movement patterns. For instance, the current LD does
not capture the case where a node jumps between its original position
and a second position with a frequency that is not significant in
terms of the potential delay it causes. Such movement will trigger
repeated recomputation, which brings in more delay than if such frequent
hopping would simply be disregarded. In as far as mobility patterns
are concerned, LD captures link stability of nodes that do not reach
their link break threshold. However, cannot distinguish between a
temporal and permanent link break.

\subsection{Node Degree Stability/Rate of Changing Neighbors\label{sub:Node-Degree-Stability/Rate of change of neighbours}}

The \emph{node degree} $N_{i}$ of a node i corresponds to the number
of neighbors $i$ has at a particular instant in time. From a mobility
perspective, an increase in node degree either means that other nodes
moved towards node $i$, or that node $i$ moved towards other nodes.
It should be noticed that from our mobility analysis perspective,
having nodes moving towards others is the same as having static nodes
simply joining or leaving a network. Hence $ND_{i}$ per se is not
an adequate mobility tracking parameter. However, if one considers
the variation of the node degree through time one may be able to infer
some mobility properties. We define \emph{Node Degree Stability} ($NDS_{i})$
for node $i$ as a parameter that tracks the rate of ND changes. For
the sake of clarity we provide a simplistic and initial embodiment
of $NDS_{i}$ in (\ref{eq:NDS1}) which corresponds to the difference
between the ND at a previous instant $t-1$ and the ND at the current
instant $t.$

\label{eq:NDS}
\begin{equation}
NDS_{i}=ND_{i_{t-1}}-ND_{i_{t}}\label{eq:NDS1}
\end{equation}

Let us consider the case where node $i$ is moving through clusters
of other nodes. $NDS_{i}$ captures, through time, the fluctuation
of neighborhood variation from the perspective of node $i$. Per se,
it does not suffice to truly track mobility, given that $i$ may be
moving or instead, $i$ may actually be static and its neighbors may
move. As far as mobility patterns are concerned, because rate change
of neighbor is not expected to change in a topology where most node
movements are confined, it is expected that that in itself will allow
the parameter to capture stability in the links of confined topology.
Worthy of mention is that the parameter in the confined node mobility
is not affected by frequency of confined motion and carters for both
long and short links as there is generally constant neighboring nodes.
However, the performance in non confined node mobility is dependent
on the gradient of change of neighbor topology and how fast the node
is moving relative to other nodes whether it is the same set of neighbors
or not. For example, a node with circular motion response to mobility
depends on the how fast it is moving relative to other nodes, when
it can be moving within the same set of nodes. It is only in confined
mobility that the types of links do not matter. In case of ping pong
and free movements, long links are affected more than short links
as long links are close to changing neighbor hood than short links
with minimal displacement. Also relative node speed, frequency and
node density plays a part and an increase of any brings about more
neighbor change signaling high mobility. With this, rate of change
of neighbors does not distinguish between temporal breaks and permanent.

\subsection{Ratio of Static vs. Moving Nodes}

The ratio of static vs. moving nodes (or a ratio between them) in
a network estimated through time and for the perspective of a single
node $i$ is here defined to be an evolution of $NDS_{i}$ and a parameter
that can be considered in order to partially capture mobility dynamics
of a network. Through time, if the percentage of nodes moving is low
in comparison to the static nodes, it is more likely to have more
stable links. Relevant also is to be able to capture the dispersion
(and not only the percentage) of such nodes on the network.

\subsection{Average Number of Link Breaks}

Another parameter that assists in tracking mobility dynamics is the
\emph{Average Number of Link Breaks} estimated in a specific interval
for a node i, $ALB_{i}$ \cite{liangLD}. If node $i$ experiences
a high ALB, then through time it may be a node to avoid, if the goal
is to provide robust paths. This implies that it Albeit interesting
due to the easy computation of such parameter, ALB can only assist
in terms of the route discovery phase, given that it may assist in
setting up more robust paths. However, and for the case of repetitive
movement patterns, ALB cannot capture such mobility dynamics and will
result in route re-computation. This is a parameter that captures
the stability of short links but not on long links in any mobility
pattern. It fails to capture even spatial correlation of long links
with confined mobility pattern. It is affected by high frequency of
repetitive node motion.

\subsection{Pause Time}

\emph{Pause time} is the period of time that the node is \emph{stationary}
(e.g. its speed is zero) \cite{Tsumochi_pausetime}. Khamayseh et
al. \cite{Khamayseh_Pausetime} have used pause time to determine
mobility levels, by assuming that nodes with long pause times are
less mobile than nodes with small pause times, and hence assist in
developing more stable links. Their notion of pause time is based
on an aggregate perspective and relative to the global time of a simulation.
In real-time, nodes exhibiting short-pause times may or may not prevent
the development of more robust links, but this is related not only
to being static, but also to the movement pattern they exhibit.

\subsection{Summary}

Most of the mobility parameters previously discussed in this section
are able to partially capture mobility dynamics of a node. Out of
the ones described, LD seems to be the most relevant to consider in
regards to attempting to develop routing metrics that can assist in
tracking mobility dynamics in particular regarding node movement with
patterns that exhibit some recurrent behavior (e.g. ping-pong movement).
The remainder parameters are relevant and may be applied to assist
parameters such as LD, in building more robust metrics.

It is, however, our belief, that LD requires a more thorough characterization
to be able to integrate routing metrics that are more sensitive to
movement.

\section{Making Routing Mobility Aware, Link Duration Heuristics\label{sec:Making-Routing-Mobility}}

In this section, we provide our proposal concerning heuristics that
augment the current definition of LD in a way that optimizes the routing
process, by being able to better capture when route re-computation
is truly required.

\subsection{Link Duration based on Signal-to-Noise Ratio Threshold}

Today's LD definition simply relates to the receiver's perceived signal
strength, as mentioned in section \ref{sub:Parameters}. As long as
two nodes are within each other transmission range, then the LD keeps
increasing \cite{liangLD}. If two nodes $i$ and $j$ share a short
link and exhibit a synchronous mobility pattern thus keeping the same
average distance while moving, then LD also increases. Let us now
consider the same type of movement pattern between two nodes that
are at a larger distance. Because of displacements nodes in the long
link will lead to reach the breaking threshold. The resultant link
duration will be much smaller than in a short link.Hence, micro-movement
of nodes will impact longer links more than shorter links in what
concerns route re-computation. However, metrics that just consider
LD cannot capture this behavior. 

The current definition of LD relates to applying a threshold for SNR,
so that LD becomes dependent of a ``good'' SNR level, that can assist
in reaching a better trade-off in terms of route re-computation vs.
signaling overhead. In other words, LD would become the instant in
time that starts when two nodes encounter each other with SNR above
the required threshold and which ends when the perceived SNR at the
receiver is lower than a pre-defined threshold. This definition is
evidently highly dependent on the choice of the SNR threshold. 

Our belief is that by working the LD definition, one can provide a
routing metric that is more adequate to environments where nodes are
expected to frequently move and hence, where there is a high variability
in terms of topology reconfiguration. In the next sections we propose
a few LD variations and explain how they may assist routing in providing
more sensitivity to node movement.

\subsection{Relaxing the Link Duration Definition: LD with a Tolerance Interval}

One way to assist routing in becoming more tolerant to frequent and
temporary movement, the LD definition can be relaxed in regards to
time. Hence we propose the \emph{Relaxed Link Duration (RLD)} heuristic.
Let us assume that a link has been stable since instant $t_{0}$ and
breaks at an instant $t_{1}$ and that at instant $t_{2}$the node
that originated such break returns to its original position. With
the current LD definition, the link would be considered broken at
instant $t_{1}$ thus originating a route re-computation. In such
case, $LD=t_{1}-t_{0}$. At instant $t_{2}$, the route would again
have to be recomputed. Following the same example provided, at instant
$t_{1}$, the node detecting the link break would wait $\Delta t$
time units before issuing a route re-computation. Assuming that $\Delta t$
is large enough (for our example, larger than $t_{2}-t_{1}$), then
the node would return, on instant $t_{2},$ to its original position
and the routing process would disregards the temporary link break.
The key aspect here is the choice of an adequate $\Delta t$. If $\Delta t$
is too large, then for nodes that may cross frequently their original
position, link breaks will not be detected and hence signaling is
reduced. The flip-side to this is that for nodes that exhibit movement
that result in a permanent link break, there is an additional delay
added.

\subsection{Spatial Stability-based Link Duration}

In section \ref{sub:Parameters} we have introduced $NDS_{i}$ and
$ALB_{i}$. The combination of $NDS_{i}$ and of $ALB_{i}$ assists
in capturing some properties of node mobility. To provide an example
of how this can be achieved, we define \emph{Spatial Stability}-\emph{based
Link Duration (SSLD)} as an extension of LD based on a correlation
between $ALB_{i}$ and $NDS_{i}$. It should be noticed that the work
here provided is intended to be initial and hence we do not provide
a concrete instantiation of a formula that may represent SSLD. Instead,
we explain the rationale for this heuristic.

Let us consider a node $i$ that has a high NDS and a high ALB. In
terms of node mobility and its impact on routing, links related to
such node are expected to be less robust, given that there is a strong
movement associated to the perspective of such node. Hence what SSLD
can provide is a way to, at an early instant in time, discard successors
of a node because they exhibit some mobility variability. As mentioned
before, the node may be static and yet, the result are less stable
paths (because most neighbors may be moving). For a more detailed
explanation we provide in Table \ref{tab:NDS-and-ALB} the full set
of NDS vs. ALB, by considering a high and low value. For each combination
we provide examples of scenarios that may result in such values. Moreover,
we also explain what the correlation may provide for each combination.

\begin{table*}[t]
\protect\caption{NDS and ALB correlation for node $i$\label{tab:NDS-and-ALB}.}

\begin{tabular}{|>{\centering}p{2cm}|>{\raggedright}p{3cm}|>{\raggedright}p{4cm}|>{\raggedright}p{4cm}|l|}
\hline 
{\small{}$NDS_{i}$} & {\small{}Low} & {\small{}Low} & {\small{}High} & {\small{}High}\tabularnewline
\hline 
{\small{}$ALB_{i}$} & {\small{}Low} & {\small{}High} & {\small{}Low} & {\small{}High}\tabularnewline
\hline 
\hline 
{\small{}Examples resulting in such parameter values} & \multicolumn{1}{>{\centering}p{3cm}|}{{\small{}\uline{Case 1}}{\small \par}

{\small{}A static node i has only a few mobile neighbors.}\linebreak{}
{\small \par}

{\small{}\uline{Case 2}}{\small \par}

{\small{}Node and its neighbors are moving exhibiting a group movement
pattern.}} & \multicolumn{1}{>{\raggedright}p{4cm}|}{{\small{}\uline{Case 3}}{\small \par}

{\small{}A static node i has a few mobile nodes which either exhibit
ping-pong behavior or which are on the move.}\linebreak{}
{\small \par}

{\small{}\uline{Case4:}}{\small \par}

{\small{}A node i keeps a stable rate of change of neighbors based
on different neighbors (and hence different links).}} & \multicolumn{1}{>{\raggedright}p{3cm}|}{{\small{}\uline{Case 5}}{\small \par}

{\small{}Node i is static and experiences a significant change in
terms of neighborhood.}\linebreak{}
{\small \par}

{\small{}\uline{Case 6}}{\small \par}

{\small{}Node i is mobile and experiences an increase in its neighbor
degree, thus implying that there is a group movement pattern - old
links are kept.}} & \multicolumn{1}{>{\raggedright}p{3cm}|}{{\small{}\uline{Case 7}}{\small \par}

{\small{}A node moves towards a more dense region. }{\small \par}

{\small{}\uline{Case 8}}{\small \par}

{\small{}A node experiences a significant change in its dense neighborhood.}}\tabularnewline
\hline 
{\small{}What the correlation assists in capturing} & {\small{}- Realizes that there is low mobility impact - - Ensures
robustness of node as potential successor} & \multicolumn{1}{>{\raggedright}p{3cm}|}{{\small{}- Realizes that there is significant mobility impact. }{\small \par}

{\small{}- Node may be ruled out as potential successor.}} & \multicolumn{1}{>{\raggedright}p{3cm}|}{{\small{}- Indication of group movement pattern; }{\small \par}

{\small{}- Stability in the neighborhood independently of having the
node static or moving.}} & \multicolumn{1}{>{\raggedright}p{3cm}|}{{\small{}- High variability in terms of movement and of neighborhood.}}\tabularnewline
\hline 
\end{tabular} 
\end{table*}

Scenarios where a node $i$ holds both a low NDS (be it negative or
positive) and low ALB imply that despite the reduction on the number
of neighbors, the rate of link breaks is low. Potential cases for
such result are scenarios where $i$ is static and has a few nodes
around it moving, or the node and its neighbors exhibit a group movement
pattern. In both cases, the correlation between $NDS_{i}$ and $ALB_{i}$
can capture that mobility is not having particular impact on routing,
for links associated to $i$.

Let us now consider a scenario with a low $NDS$ and a high $ALB$
value. This means that despite the fact that $i$ has stability in
terms of neighbors (i.e. it keeps a steady rate between neighbors
leaving and joining), it can be captured that the neighborhood of
$i$ shows high variability. In contrast, if only $NDS$ would be
considered, then it would be assumed that $i$ was a node stable in
terms of links and hence a robust choice for a successor on a path.
Also, if only $ALB$ would be considered on such a scenario, then
it would be immediately assumed that $i$ was a node to rule out. 

As example of scenarios which result in a high value for $NDS$ and
low value for $ALB$ we provide cases 5 and 6. For both cases $i$
sees an increase in neighbors (new nodes are joining); yet, such change
has no impact in terms of links already established. If $i$ is a
static node (cf. case 5) then this means that its neighborhood is
mobile in its majority. If instead $i$ is moving, there is a clear
group movement pattern, given that there is no increase in terms of
link breaks.

The final discussion here provided relates to having both a high $NDS$
and a high $ALB$. Potential examples of scenarios where this occurs
are cases 7 and 8 are where node \emph{$i$} is respectively experiencing
a sudden and significant change in terms of neighborhood or moving
through a dense region. 

In the next section we provide a brief analysis related to the performance
comparison for the heuristics described in this section.

\section{Performance Comparison Analysis}

\label{sec:Analysis-1}

This section provides an analysis on the performance comparison for
the heuristics proposed on section \ref{sec:Making-Routing-Mobility}.
We consider a specific set of scenarios and evaluate the different
heuristics in terms of route re-computation in the form of signaling
overhead reduction, delay reduction, as well as improved throughput.
Also looked into is the expected outcome as far as different routing
families will be affected by the different heuristics in their respective
routing phases(i.e routing discovery and routing maintenance). As
underlying scenario for this comparison we consider Figure \ref{fig:Node-Mobility-and-Routing-1}
main cases: i) confined movement to a specific region where node a
is making confined movements; ii) ping-pong movement as a result of
periodic node mobility of node b; iii) free movement in a large region.
The confined movement example stands for a case where the node exhibits
some features that may be captured, as happens in social mobility
models, but its mobility is confined to a specific region and hence
there may be some probability of returning to origin. In this case
it is also likely that most of the node's movements do not truly impact
the links it is part of severely. In other words, movement does not
always imply a link break. The second case stands for an example of
movement of a node which returns to the original position frequently.
The third case represents an example of movement where the node will
not return to origin. 

In addition to the movement patterns based on Figure \ref{fig:Node-Mobility-and-Routing-1},
we also consider other parameters in our analysis. From a topology
perspective we shall analyze the different cases assuming different
distances for links, namely, short and long links, given that as explained,
movement of nodes impacts differently the network depending on the
distance of nodes to neighbors. We also consider \emph{network density},
namely, low or high density. A third parameter that we take into consideration
is \emph{movement frequency} in the form of \emph{low frequency} (the
node moves seldom thus exhibits long pause times) and \emph{high frequency}
(the node is highly dynamic; pause time is short). We have also analyzed
the impact of routing based on the routing family; Link State or Distance
Vector highlighting the difference in impact of mobility on routing
based on different LD heuristics, covering also the impact on the
routing processes( i.e. route discovery and route maintenance).

In Table \ref{tab: Comparison}, we provide a summarized comparison
of the performance of each heuristic, namely, LD based on today's
definition (benchmark), and the two proposed heuristics: LD with a
tolerance interval, as well as LD based on Spatial Stability.

The table contemplates the three scenarios (confined movement category,
ping-pong movement, free movement) where we have considered the combination
of network density and movement frequency by relying on the two extreme
values \emph{low} and \emph{high}. For each scenario, we also consider
what may happen if the topology is mostly based on short or long links. 

\begin{table*}[tbh]
\protect\caption{Comparison of the LD heuristics.\label{tab: Comparison}}

\begin{tabular}{|>{\centering}p{0.9cm}|>{\centering}p{0.8cm}|>{\centering}p{1cm}|>{\raggedright}p{2cm}|>{\raggedright}p{2cm}|>{\raggedright}p{2cm}|>{\raggedright}p{2cm}|>{\raggedright}p{2cm}|>{\raggedright}p{2cm}|}
\hline 
\multicolumn{1}{|>{\centering}p{0.9cm}}{} & \multicolumn{1}{>{\centering}p{0.8cm}}{} &  & \multicolumn{2}{>{\centering}p{2cm}|}{\noun{\tiny{}Confined Movement}} & \multicolumn{2}{c|}{\noun{\tiny{}Ping-pong }} & \multicolumn{2}{c|}{\noun{\tiny{}Free Movement}}\tabularnewline
\hline 
\textbf{\scriptsize{}Heuristic} & \textbf{\scriptsize{}Network density} & \textbf{\scriptsize{}Movement frequency} & \textbf{\scriptsize{}Short links} & \textbf{\scriptsize{}Long links} & \textbf{\scriptsize{}Short links} & \textbf{\scriptsize{}Long links} & \textbf{\scriptsize{}Short links} & \textbf{\scriptsize{}Long links}\tabularnewline
\hline 
\hline 
\textbf{\scriptsize{}LD } & {\tiny{}Low} & {\tiny{}Low} & {\tiny{}- Path re-computation hardly triggered without need} & {\tiny{}- Path re-computation hardly triggered without need but error
probability higher than with short-links} & {\tiny{}-Some path re-computation}{\tiny \par}

{\tiny{}-Some throughput reduction and delay increase.}{\tiny \par}

{\tiny{}- only affects parts of the network} & {\tiny{}-Some path re-computation}{\tiny \par}

{\tiny{}-Some throughput reduction and delay increase}{\tiny \par}

{\tiny{}- Worse performance overall than in short-link} & {\tiny{}- Path re-computation triggered without need if node speed
is low.} & {\tiny{}- Path re-computation triggered without need if node speed
is low.}\tabularnewline
\cline{2-9} 
 & {\tiny{}Low} & {\tiny{}High} & {\tiny{}- Path re-computation triggered without need}{\tiny \par}

{\tiny{}- Movement frequency dictates performance degradation} & {\tiny{}- Path re-computation triggered without need}{\tiny \par}

{\tiny{}- Movement frequency dictates performance degradation} & {\tiny{}- Path re-computation triggered without need}{\tiny \par}

{\tiny{}- Movement frequency dictates performance degradation}{\tiny \par}

{\tiny{}- only affects parts of the network} & {\tiny{}- Path re-computation triggered without need}{\tiny \par}

{\tiny{}- Movement frequency dictates performance degradation}{\tiny \par}

{\tiny{}- Worse performance overall than in short-link} & {\tiny{}- Path re-computation triggered without need if node speed
is low.}{\tiny \par}

{\tiny{}- Link break hard to predict/anticipate} & {\tiny{}- Path re-computation triggered without need if node speed
is low.}{\tiny \par}

{\tiny{}- Link break hard to predict/anticipate}\tabularnewline
\cline{2-9} 
 & {\tiny{}High} & {\tiny{}Low} & {\tiny{}- Path re-computation hardly triggered without need} & {\tiny{}-Path re-computation hardly triggered without need}{\tiny \par}

{\tiny{}- Slightly worse performance overall than in short-link} & {\tiny{}-Some path re-computation} & {\tiny{}-Some path re-computation}{\tiny \par}

{\tiny{}- Worse performance overall than in short-link} & {\tiny{}- Path re-computation triggered without need if node speed
is low.}{\tiny \par}

{\tiny{}- Plenty of alternative paths} & {\tiny{}- Path re-computation triggered without need if node speed
is low.}{\tiny \par}

{\tiny{}- Link break hard to predict/anticipate}\tabularnewline
\cline{2-9} 
 & {\tiny{}High} & {\tiny{}High} & {\tiny{}- Path re-computation hardly triggered without need}{\tiny \par}

{\tiny{}-Throughput reduction and delay increase due to MAC contention} & {\tiny{}-Heavy path re-computation}{\tiny \par}

{\tiny{}-Heavy throughput reduction and delay increase due to MAC
contention}{\tiny \par}

{\tiny{}- Worse performance overall than in short-link} & {\tiny{}-Heavy path re-computation}{\tiny \par}

{\tiny{}-Heavy throughput reduction and delay increase due to MAC
contention} & {\tiny{}-Heavy path re-computation}{\tiny \par}

{\tiny{}-Heavy throughput reduction and delay increase due to MAC
contention}{\tiny \par}

{\tiny{}- Worse performance overall than in short-link} & {\tiny{}- Path re-computation triggered without need if node speed
is low.}{\tiny \par}

{\tiny{}-Plenty of alternative paths} & {\tiny{}- Path re-computation triggered without need if node speed
is low.}{\tiny \par}

{\tiny{}- Link break hard to predict/anticipate}\tabularnewline
\hline 
\hline 
\textbf{\scriptsize{}RLD} & {\tiny{}Low} & {\tiny{}Low} & {\tiny{}- Similar performance to LD but slight increase in delay} & {\tiny{}- Similar performance to LD but slight increase in delay} & {\tiny{}- slight improvement when compared to LD} & {\tiny{}- slight improvement when compared to LD} & {\tiny{}- Slightly worse performance than LD} & {\tiny{}- Slightly worse performance than LD}\tabularnewline
\cline{2-9} 
 & {\tiny{}Low} & {\tiny{}High} & {\tiny{}- Good improvement compared to LD}{\tiny \par}

{\tiny{}- Less sensitivity to movement frequency } & {\tiny{}- Good improvement compared to LD}{\tiny \par}

{\tiny{}- Less sensitivity to movement frequency } & {\tiny{}- Good improvement compared to LD} & {\tiny{}- Good improvement compared to LD}{\tiny \par}

{\tiny{}- Worse performance overall than in short-link} & {\tiny{}-Good improvement to LD if node speed low and node pause time
is high and frequent} & {\tiny{}-Good improvement to LD if node speed low and node pause time
is high and frequent}\tabularnewline
\cline{2-9} 
 & {\tiny{}High} & {\tiny{}Low} & {\tiny{}- Good improvement compared to LD but slight increase in delay} & {\tiny{}- Good improvement compared to LD but slight increase in delay} & {\tiny{}- Good improvement compared to LD} & {\tiny{}- Good improvement compared to LD}{\tiny \par}

{\tiny{}- Worse performance overall than in short-link} & {\tiny{}- Similar to LD} & {\tiny{}- Similar to LD}\tabularnewline
\hline 
 & {\tiny{}High} & {\tiny{}High} & {\tiny{}- Good improvement compared to LD}{\tiny \par}

{\tiny{}- Less sensitivity to movement frequency } & {\tiny{}- Good improvement compared to LD}{\tiny \par}

{\tiny{}- Better sensitivity to movement frequency } & {\tiny{}- Good improvement compared to LD} & {\tiny{}- Good improvement compared to LD}{\tiny \par}

{\tiny{}- Worse performance overall than in short-link} & {\tiny{}-Good improvement to LD if node speed low and node pause time
is high and frequent} & {\tiny{}-Good improvement to LD if node speed low and node pause time
is high and frequent}\tabularnewline
\hline 
\hline 
\textbf{\scriptsize{}SSLD} & {\tiny{}Low} & {\tiny{}Low} & {\tiny{}- Similar performance to LD} & {\tiny{}- Similar performance to LD} & {\tiny{}- Similar performance to LD} & {\tiny{}- Similar performance to LD} & {\tiny{}- Similar performance to LD} & {\tiny{}- Similar performance to LD}\tabularnewline
\cline{2-9} 
 & {\tiny{}Low} & {\tiny{}High} & {\tiny{}- Slight improvement compared to LD}{\tiny \par}

{\tiny{}- No delay added} & {\tiny{}- Slight improvement compared to LD}{\tiny \par}

{\tiny{}-No delay added} & {\tiny{}- Good improvement compared to LD} & {\tiny{}- Good improvement compared to LD} & {\tiny{}- Slightly better performance than LD } & {\tiny{}- Slightly better performance than LD}\tabularnewline
\cline{2-9} 
 & {\tiny{}High} & {\tiny{}Low} & {\tiny{}- Slight improvement compared to LD}{\tiny \par}

{\tiny{}- No delay added} & {\tiny{}- Slight improvement compared to LD}{\tiny \par}

{\tiny{}- No delay added} & {\tiny{}- Slight improvement compared to LD}{\tiny \par}

{\tiny{}- No delay added} & {\tiny{}- Slight improvement compared to LD}{\tiny \par}

{\tiny{}- No delay added} & {\tiny{}- Slightly better performance than LD} & {\tiny{}- Slightly better performance than LD}\tabularnewline
\hline 
 & {\tiny{}High} & {\tiny{}High} & {\tiny{}- Good improvement compared to LD}{\tiny \par}

{\tiny{}- No delay added} & {\tiny{}- Good improvement compared to LD}{\tiny \par}

{\tiny{}- No delay added} & {\tiny{}Slight improvement compared to LD}{\tiny \par}

{\tiny{}- No delay added} & {\tiny{}- Slight improvement compared to LD}{\tiny \par}

{\tiny{}- No delay added} & {\tiny{}- Slightly better performance than LD} & {\tiny{}- Slightly better performance than LD}\tabularnewline
\hline 
\end{tabular}
\end{table*}

\subsection{Confined Movement Scenario}

We start by discussing the performance of the heuristics in confined
movement scenarios where the majority of links are short. Being a
confined movement scenario, the nodes involved move in a specific
personal space revolving around their initial position and hence are
expected to return or to pass on its original position with some frequency.
Such frequency is strongly related to the link stability. 

Because confined movements of nodes result in few permanent link breaks
for short links, if any, confined movement has almost no impact on
routing performance. Such mobility, even though, will result short
links elongating to long ones, no notable link breaks will occur,
if any, then few. With few links breaks and absence of major change
in the node neighborhood, the three heuristics will give almost the
same result.

For the long links, because a slight node displacement cause links
to reach the breaking threshold, varied performance is expected under
the three heuristics. We look into the distance vector family and
the respective heuristics at the different stages of routing. In a
low node topology region with low node mobility frequency, the route
discovery of distance vector under LD will record few link breaks
and it is expected that node mobility will be at its minimal. As such,
route discovery will barely be affected as it takes a short period.
However, there exist a possibility of having unutilised paths as nodes
may be resting at some places resulting in discovery of long paths.
In the route maintenance phase, because few links are expected to
break, a slight increase in path recomputation will occur under LD.
As far as RLD and SSLD are concerned, route discovery latency is expected
to be the same as LD due to few breaking links and nodes resting in
same positions for long periods. In the route maintenance phase, RLD
and SSLD will slightly outperform LD, due to the few path recomputations
that will occur. With increase in node mobility frequency and in a
low density topology, more long links in the face of confined node
mobility will be reaching the breaking threshold. Compared to the
static and nodes with spatial correlation, Distance vector routing,
using LD heuristic, will be challenged even at a routing discovery
level. This is due to the fact that propagation of the control messages
will not be effective resulting in high latency due to link breaks.
The control messages, as a result of failed discovery and quests to
discover new paths, will contribute to the delay also. As far as route
maintenance is concerned, there will be more path recomputation compared
to low mobility frequency and in short links. The control overhead
will weigh down the throughput and increase the delay. If RLD is used
on the other hand, route discovery latency will be low as route discovery
control messages will propagate in the ``tentative'' links also.
An improved performance is expected in the route maintenance also,
as less path recomputations will occur. With SSLD, the avoidance of
nodes with high mobility will bring a slight improvement in route
discovery and maintenance phase of the distance vector.

A topology with high node degree gives a different perspective of
mobility impact different from the low topology. Still looking at
long links, low confined node mobility will not impact routing discovery
in that more alternative paths exist and that possibility of having
good routes is high. The challenge is that periodic neighbor control
messages will increase and route discovery will record the increase
in delay. Because the delay is not as a result of change in topology,
the three heuristic performance will be the same as far as route discovery
in distance vector will be the same. Routing mainteinance performance
will be the same, with LD giving slightly more delay to due few link
breaks. When long links in a high node topology are subjected to high
frequency confined node mobility, a number of factors come into play,
MAC contention and link breaks become common. With this, the different
heuristics will render the distance vector routing protocol different
performance levels. Route discovery using LD heuristic will affected
due to MAC contentions and failed route discoveries as a result of
link breaks increasing the delay. Under RLD, the delay will only be
due to MAC contentions as control messages will propagate the tentative
links too. SSLD will have better performance as LD in that the stable
nodes will be fished out. In as far as routing maintenance is concerned,
RLD will have a good improvement as some temporal link breaks will
not cause path recomputations and SSLD will capture stable nodes for
its routing purposes.

We now look into how confined node mobility will affect a link state
routing protocol. Because of the proactive nature of the routing family,
paths are computed with change in topology ( i.e. link state, neighbor
list). As such, the impact mobility of such a protocol, in as far
as routing data is concerned, will be in the maintenance phase. We
use the same parameters as in distance vector to analyze the impact
on link state. These being; short or long links, low or high node
mobility frequency and low or high node degree. As mentioned above
on short links, confined node mobility will lead to almost no link
breaks and the three heuristics will behave the same, with expected
increase in delay in high node degree topologies. Because no extra
control packets are generated, routing degradation due to mobility
may be due to different reasons at some instances when compared to
distance vector.

In a low node degree topology, low confined node movement frequency
will cause the few links to break and this will cause path recomputations
when LD is used, but because these are few link breaks, few packets
will be delayed and it will have less bearing on routing in a link
state. When the mobility frequency increases, more link breaks occur
and path recomputations will occur within a limited number of nodes.
The result will be increased delay as packets are buffered as they
await the link break to be detected and path to be recomputed. When
RLD is used, expected is that temporal link breaks will detected and
the packets will propagate ``tentative'' links, hence, less packets
will be buffered and improvement is expected compared to LD. Because
there is a larger number of link breaks, the only benefit SSLD will
bring is to detect the few stable links amid highly mobile nodes and
the improvement is slight. An increase in neighbor degree will cause
more control packets,\textbf{ }with highly mobile nodes, a good number
of links will reach the breaking threshold and path recomputation
will be common with LD as a heuristic. The extra control overhead
will weigh down the performance. Unlike in low density topology where
few available paths break, a number of paths will be available but
they are short lived. RLD will also be weighed down due to control
overhead. Slight improvement is expected in SSLD. Low node mobility
in a high node density topology will introduce overhead to routing
and all heuristics will be affected with RLD and SSLD with better
delay than LD as few path recomputations do not occur.

In summary, a low density network where most nodes are static (movement
frequency is low), as shown in the first line of table \ref{tab: Comparison},
then the expected result when relying on LD is that the path re-computation
is normally triggered due to permanent link breaks. With the increase
of movement frequency (cf. Table \ref{tab: Comparison} second and
fourth row) , path re-computation occurs due to link breaks. A first
aspect to highlight is that the LD performance in terms of routing
degradation is heavily dependent on movement frequency, and pattern.
A second aspect relates to the fact that LD behavior will impact routing
more heavily for dense networks. Overall, LD performance is dependent
on the link distance and hence the impact on routing is expected to
be more significant for topologies where links are in their majority
long.

Looking into the first proposed heuristic, RLD (cf. Table \ref{tab: Comparison}
rows 5 to 8), for the case of confined movement. At a first glance
and in what concerns simpler topologies (low density, mostly static
nodes) RLD seems to have a similar performance at the cost of a slight
increase in delay, which depends on the choice of the tolerance interval.
It should be noticed that RLD will experience an additional delay,
but when LD is applied, there is also a delay increase due to path
re-computation. Therefore, even though there is potentially an increase
in delay depending on the choice of the tolerance interval, we expect
RLD to behave better given that it reduces the need to recompute paths
and hence improves throughput and reduces the delay associated to
path re-computation. Such improvement is not so significant in simpler
topologies but becomes significant when the movement frequency increases,
given that the RLD is less sensitive than LD to movement frequency.

In regards to SSLD, for simpler topologies the expected performance
is similar to LD, for this scenario. The reason for this is that SSLD
does not incur particular delay (in contrast to RLD) and the variability
in terms of ND and of link breaks is low. When the movement frequency
increases, then this heuristic works better by being able to more
quickly track that there is variability in terms of movement affecting
the link.

\subsection{Ping-Pong Scenario}

Ping pong effect, unlike confined node mobility where impact on routing
highly depends on the link length, affect both short and long links.
This is in a situation where relative node displacement can be so
large such that short links will elongate and even reach the link
break threshold. We discuss the impact the ping pong effect will have
in routing , extending the analysis to the routing processes under
varied conditions of node degree and mobility frequency.

Ping pong effect, as mentioned earlier, can result in short links
becoming long with the worst case having links reaching their breaking
threshold. In a low node density topologies with minimal node mobility,
expected is that few links will break both short and long links with
short links being tolerant to displacement. As such, route discovery
will barely be impact in that node will rest in their respective positions
for some time. The three different heuristics are expected to perform
the same as there is barely changes in link breaks and node neighborhood
with LD with slight delay as there is a probability of picking on
the node that is in motion for route establishment. As for route maintenance
phase, expected is that LD and SSLD will perform the same, while RLD
will introduce slight delay as the few links that have broken maybe
held on to in the tolerance interval. On the other hand, during the
routing maintenance phase, RLD is expected to give the best performance
due to reduced path recomputation even though they are already few.
An increase in mobility frequency, increases the number of link reaching
threshold and Routing maintenance best performance is expected to
be from RLD as there are less control overhead. SSLD will record a
slight improvements due to selection of stable links. LD, on the other
hand, will be affected due to increased routing overhead in a sparsely
populated topology with few alternative paths for routing. The route
discovery using the RLD will have less latency as links picked are
bound to successfully deliver the route discovery control messages
even in their ``tentative'' state. LD and SSLD will have the same
performance. Increased node density have pros and cons. Increased
or high node density provides a number of alternative paths in case
of link breaks but bring along also control overhead. With increased
mobility, RLD, once more will give the best performance in both routing
maintenance and discovery. When node mobility frequency is low, the
performance of the heuristics will almost be the same as few links
will break. Compared to low node density and low mobility frequency,
in high node density and low mobility nodes, the heuristics are expected
to give a scaled down performance with increase in node degree due
to MAC contention.

In summary, ping pong effect is a scenario where LD does not suffice
to provide routing with adequate stability in the face of movement routing.
The ping-pong frequency is highly related to the link stability. If
it is too short (e.g. a few milliseconds) it will originate frequent
link breaks. If it is long, then it affects routing less, when considering
LD. Hence, when LD is considered this type of movement results in
significant throughput reduction and increase in delay, as well as
additional signaling due to the route re-computation.

By considering RLD it is possible to decrease the impact of mobility
on routing, if the tolerance interval is adequately tuned. The result
is an increase in performance due to lesser path re-computation. Another
aspect that seems interesting to be further explored in RLD is that
it is an heuristic that seems to be less sensitive than LD to movement
frequency.

SSLD is also expected to offer a good improvement in comparison to
LD, for this scenario. The correlation between NDS and ALB gives the
means to detect that there is some repetition in terms of neighborhood.
When combined to the linear ALB, it is feasible to realize that the
node is moving in a repetitive pattern and hence to take a more intelligence
action in terms of the decision to recompute.

When comparing the three heuristics in this scenario for short links
vs. long links, we believe that the behavior is similar to the previous
scenario: there is an overall performance degradation for the case
of long links, given that distance increases the need for MAC contention.

\subsection{Free Movement Scenario}

In the free movement scenario, we consider the case where a node is
moving away from its original position and not expected to return.
The way it moves away (pattern) also impacts the movement. In this
type of situation, LD suffices to ensure that path re-computation
is only triggered when a permanent link break occurs. When we consider
RLD, then the expected performance is similar to the one of LD if
the topology is mostly static but when the movement frequency increases
and with the increase in network density then the behavior of RLD
becomes worse as it adds delay to the moment when a link breaks. It
should be noticed that such delay, albeit always expected to be present,
can be tuned through time to become less significant. An exception
to this may be the situation where a node moves with a slow motion
and stopping on the way for long pause times. In some cases, LD may
result in path re-computation that can be avoided an instant later
in time (due to a slow motion node, pausing also for long times).

Considering SSLD, our analysis tells us that when considering simpler
topologies, this heuristic seems to behave in a similar way to LD.
But when there is an increase in the movement frequency then SSLD
may assist in providing an action on the right instant. Moreover and
in comparison to RLD, there is no significant delay added when applying
SSLD. As for the three types of scenarios, SSLD seems to be the better
suited for the confined movement scenario.

\section{Conclusions and Future Work}

\label{sec:Concl}

This paper addresses ways to make multihop routing more sensitive
to movement. We introduce the problem space of impact of mobility
on routing and discuss a few existing parameters and also some novel
parameters that may be considered to track mobility in routing. Based
on the analysis and discussion of such parameters, we propose heuristics
based on the notion of link duration to attempt to make routing more
sensitive to node mobility. Such heuristics are then compared against
the current definition of link duration for a meaningful set of scenarios.

Albeit being initial work that requires further delving, the provided
comparison hints that the two heuristics being proposed, namely, the
RLD and the SSLD, are relevant enough to be considered as potential
candidates to assist multihop routing in terms of mobility sensitivity.
While the RLD seems to be more relevant for scenarios that exhibit
some repetitive motion pattern, SSLD seems to provide an overall good
performance, in particular for confined movement scenarios.

As future work we intend to further detail the two proposed heuristics,
and to provide an evaluation of them against LD for the most popular
forms of multihop routing (distance vector and link-state approaches).

\bibliographystyle{plain}
\addcontentsline{toc}{section}{\refname}\bibliography{paper1may2010}

\begin{IEEEbiographynophoto}{}
N\textbf{amusale Chama} holds a BEng (06) in Electronic and Telecommunication
Engineering from the University of Zambia. During 2006-2008, she worked
as Radio Planning Engineer in Zain Zambia, now Airtel Zambia. Currently
she is a researcher in IANLab, University Lusofona and a PhD(MAP-Tele)
student. Her current research activities focused on the impact of
mobility in routing of Ad hoc Wireless Networks. Her research interests
are in Advance forwarding and Routing.

\vspace{3cm}

\textbf{Dr. Rute C. Sofia} graduated (95) in Informatics Engineering
from the University of Coimbra; MSc (98) and PhD (2004) in Informatics
from the University of Lisbon. She is currently the Scientific Director
for Technology in SITI, and the co-coordinator of the IAN Lab. Since
1995 she has been pursuing engineering activities in industry both
nationally and internationally. During 2000-2003 she was a visiting
scholar at the Internet Center for Advanced Internet Research (ICAIR),
Evanston, USA, and a visiting scholar at the University of Pennsylvania,
USA. Between 2004-2007 she was a senior research scientist in SIEMENS
AG Corporate Technology/Nokia-Siemens Networks, focusing on Future
Internet topics such as global mobility across multi-access networks
(e.g. Mobile IP, WiMAX, 3G) and novel forwarding paradigms (e.g. frame
routing, network coding). She was (07-10) the co-leader of the Internet
Architectures and Networking (IAN) team, UTM, INESC Porto, team which
has been recently (2010) transferred to ULHT together with the associated
activities, and integrated into the recent SITI. Her research interests
relate to advanced forwarding and routing; mobility aspects such as
management and modelling; resource management in wireless networks;
cooperative networking. She has been actively involved in IETF groups
such as the Next Steps in Signaling (NSIS), Mobility Extensions (MEXT),
V6OPS, and is an IEEE and an ACM SIGComm member.\end{IEEEbiographynophoto}

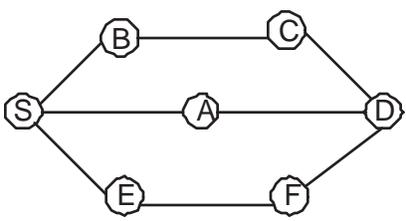
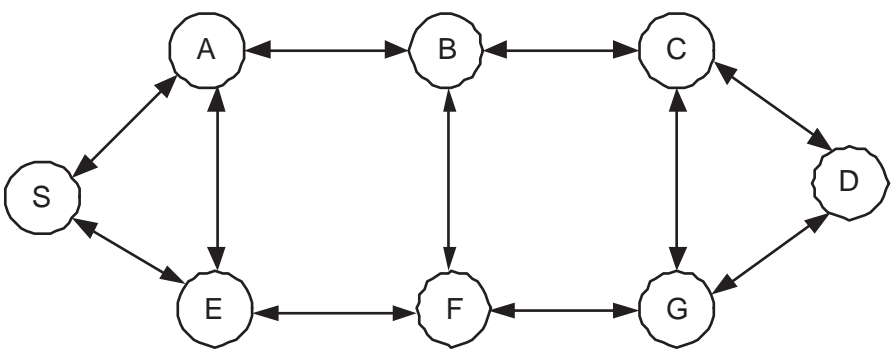
\end{document}